\def\NIMA{{\em Nucl. Instrum. Methods} A}
\documentclass[fleqn,twoside]{article}
\usepackage{espcrc2}

\usepackage{epsfig}
\usepackage{graphics}
\usepackage{graphicx}

\newcommand{\AmS}{{\protect\the\textfont2
  A\kern-.1667em\lower.5ex\hbox{M}\kern-.125emS}}

\title{The CDF MiniPlug Calorimeters at the Tevatron}

\author{K.~Goulianos, M.~Gallinaro, K.~Hatakeyama, 
S.Lami\thanks{Corresponding author. Tel.: +1-212-327-8832;
fax: +1-212-327-7786.  {\em E-mail address}: lami@fnal.gov
(S. Lami)}, C.~Mesropian, K.~Terashi \\~\\
{\em Department of Experimental Physics, The Rockefeller University,  New York, New York 10021, USA}}

\begin{document}

\begin{abstract}
Two MiniPlug calorimeters,
designed to measure the energy and lateral position of particles 
in the pseudorapidity region of $3.6<|\eta|<5.1$ of the
CDF detector, have been installed as part of the Run II CDF upgrade
at the Tevatron collider. 
Detector performance and first results from  $\bar pp$ collision data
are presented.
\vspace{1pc}
\end{abstract}

\maketitle

\section{INTRODUCTION}

The CDF diffractive physics program
relies on two MiniPlug (MP) calorimeters,
which can detect both charged and neutral particles,
to measure the event energy flow in the very forward
rapidity region on opposite sides of the interaction point.
The performance of a MP prototype is published in \cite{nim_proto},
while 
the final detector assembly and the results
from a cosmic ray test are reported in \cite{nim_final}.

\section{DETECTOR}

\begin{figure}[hb]
\vspace*{-.29in}
\epsfig{file=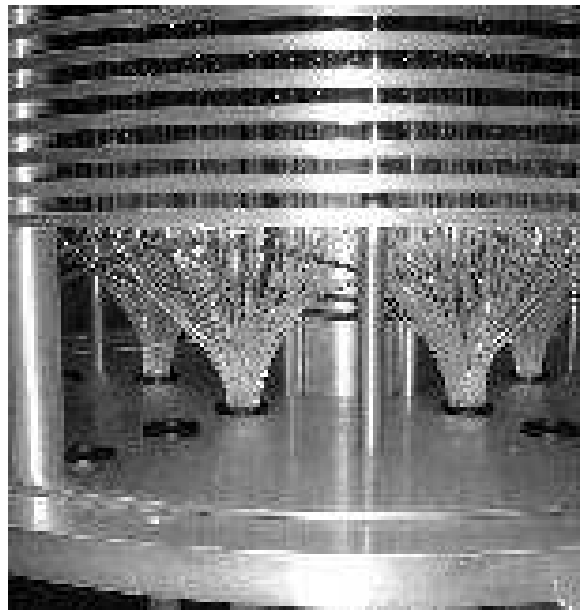,angle=0,width=18pc}
\vspace*{-.4in}
\caption{Fiber routing in the MiniPlug.}
\vspace*{-.22in}
\label{fig:routing}
\end{figure}

The MPs consist of alternating layers 
of lead plates and liquid scintillator read out by 
{\it WaveLength Shifting} (WLS) fibers (Fig.~\ref{fig:routing}).
The fibers are perpendicular to the lead plates and parallel
to the proton/antiproton beams,  and are read out 
by {\em Multi-Channel PhotoMultiplier Tubes} (MCPMTs).
The 16-channel Hamamatsu R5900 MCPMTs have a quartz window
which improves the radiation hardness.
Each MP is housed in a cylindrical steel vessel 26$''$ 
in diameter and has a 5$''$ hole concentric with  
the vessel to accommodate the beam pipe.
The depth of a MP is 32 radiation lengths and 
1.3 interaction lengths.
The short hadronic depth limits the lateral spread
of the showers. The MPs have a novel {\it towerless}
geometry with no dead regions, due to the lack of internal
 boundaries, and towers are formed by combining
the desired number of fibers. 
The centroid of the tower pulse height provides the position 
of the shower initiating particle.

The tower design is based
on an hexagon geometry. Holes in the lead plate
are conceptually grouped in hexagons, each with six holes.
The six fibers inserted in the holes of a hexagon are grouped together and are
viewed by one MCPMT channel.
 A seventh fiber, which is clear and carries
the light from a calibration LED, is also read out by each MCPMT pixel
to allow a periodical monitoring of the MCPMT response.
The MCPMT outputs are added to form 84
calorimeter towers per MP, organized in four concentric
rings around the beam pipe.
The entire MCPMT can also be read out through the
 last dynode
output to provide trigger information.
Each MP has a total of 18 trigger towers, arranged in three rings.

\section{CALIBRATION AND RUN II DATA}

Cosmic ray muons were used to test one 60$^o$ wedge of the
east MP. The cosmic ray trigger fired on a 2-fold coincidence
of scintillation counter paddles located on top and at the bottom
of the MP vessel, placed with towers pointing upward.
The single photoelectron response was measured using a randomly
gated signal from a $^{60}Co$ source.
The response to a minimum ionizing particle was found to be
above 100 photoelectrons, exceeding the design specifications.
\begin{figure}[ht]
\epsfig{file=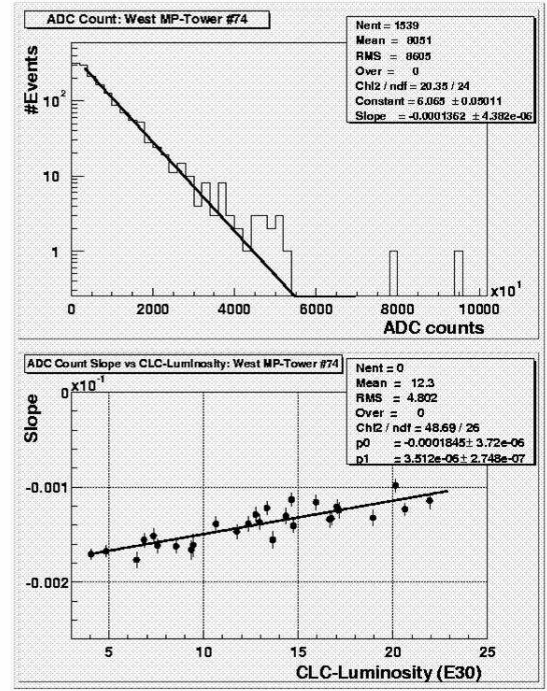,angle=0,width=18pc}
\vspace*{-.45in}
\caption{The ADC count distribution of the data can be fitted
well by a falling exponential curve ({\it top}).
The absolute slope values decrease with increasing luminosity
due to event pile-up ({\it bottom}).
Data are shown for Tower 74 of the west MP.}
\label{fig:slopes}
\end{figure}

\begin{figure}[htb]
\centering{\epsfig{file=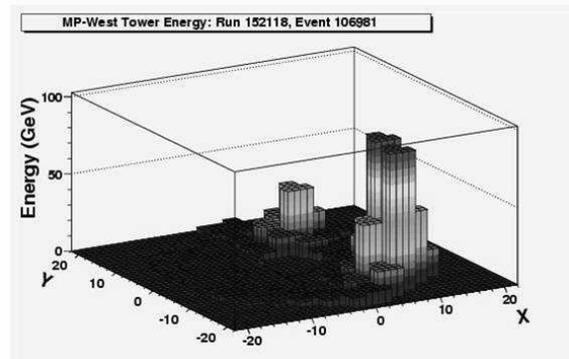,angle=0,width=18pc}}
\vspace*{-.3in}
\caption{Two hadron event  in the west MP.}
\label{fig:lego}
\end{figure}

Although a precise energy calibration is not crucial
to the analysis of diffractive processes, an attempt
was made to estimate the energy of jets and particles
using colliding beam data from Run II.
For each tower, the ADC count distribution
of a sample of minimum bias events
can be fitted
well by a falling exponential curve (Fig.~\ref{fig:slopes}, top).
A Monte Carlo simulation was then used to calibrate the pseudorapidity
dependence of the tower-by-tower response.
Due to pile-up effects at larger rapidity regions, a luminosity
dependence of the ADC count distribution slope
is observed (Fig.~\ref{fig:slopes}, bottom).
A linear fit describes well the data, with the slope
decreasing with increasing values of instantaneous luminosity.
The particle multiplicity is measured by counting clusters of towers
with energy above noise (Fig.~\ref{fig:lego}).

The MPs are presently an integral part of
the CDF detector, collecting good quality data
for further exploring the realm of diffractive physics.


\begin{thebibliography}{9}
\bibitem{nim_proto} K. Goulianos and S. Lami,
 \NIMA 430 (1999) 34.
\bibitem{nim_final} K. Goulianos {\it et al.},
 \NIMA 496 (2003) 333.
\end{thebibliography}
\end{document}